\documentclass[a4paper,journal]{IEEEtran}
\usepackage{mathrsfs}
\usepackage[dvips]{graphicx}
\usepackage{subfigure}
\usepackage{color}
\usepackage{url}
\usepackage[noadjust]{cite}
\usepackage{multicol}
\usepackage{amssymb,dsfont,amsmath,amsfonts,mathrsfs,soul}
\usepackage{xspace,colortbl,graphicx}
\usepackage{enumerate}
\usepackage[ruled]{algorithm2e}
\usepackage{color, soul, xcolor} %??color, ?? soul 
\usepackage{multirow}
\usepackage{arydshln}
\usepackage{pifont}
\usepackage{mathtools} % \coloneqq   Def.
\usepackage{cases}
\usepackage{framed}

\newtheorem{theorem}{\indent Theorem}[section]

\newtheorem{definition}{\indent Definition}[section]
\newtheorem{remark}{\indent Remark}[section]

\newcommand{\be}{\begin{equation}}
\newcommand{\ee}{\end{equation}}
\newcommand{\bea}{\begin{eqnarray}}
\newcommand{\eea}{\end{eqnarray}}
\newcommand{\ba}{\begin{array}}
\newcommand{\ea}{\end{array}}

\sethlcolor{yellow}

\ifCLASSINFOpdf

\else

\fi

\large
\begin{document}
\title{Benefits of V2V communication in connected and autonomous vehicles in the presence of delays in communicated signals 
}

\author{Guoqi~Ma, Prabhakar~R.~Pagilla$^\ast$, Swaroop~Darbha

\thanks{The authors are with the Department of Mechanical Engineering, Texas A\&M University, College Station, TX 77843, USA (e-mail: gqma@tamu.edu; ppagilla@tamu.edu; dswaroop@tamu.edu). $^\ast$Corresponding author.

Part of this work 
%has been submitted to
was presented at the 26th IEEE International Conference on Intelligent Transportation Systems (ITSC), Bilbao, Spain, 2023~\cite{ITSC2023}. 
%~\cite{ITSC2023}. 

}}

% <-this % stops a space% <-this % stops a space
% <-this % stops a space% <-this % stops a space

%\markboth{Submitted to IEEE Transactions on Intelligent Transportation Systems}%

\maketitle

%%%%%%%%%%%%%%%%%%%%%%%%%%%%%%%%%%%%%%%%%%%%%%%%%%%%%%%%%%%%%%%%%%%%%%%%%%%%%%%%%%%%%%%%%%%%%%%%%%%5
	
\begin{abstract}
 In this paper, we investigate the effect of signal delay in communicated information in connected and autonomous vehicles. In particular, we relate this delay's effect on the selection of the time headway in predecessor-follower type vehicle platooning with a constant time headway policy (CTHP).  We employ a CTHP control law for each vehicle in the platoon by considering two cases:  cooperative adaptive cruise control (CACC) strategy where information from only one predecessor vehicle is employed and CACC+ where information from multiple predecessor vehicles is employed. We investigate how the lower bound on the time headway is affected by signal transmission delay due to wireless communication. We provide a systematic approach to the derivation of the lower bound of the time headway and selection of the appropriate CTHP controller gains for predecessor acceleration, velocity error and spacing error which will ensure robust string stability of the platoon under the presence of signal delay. We corroborate the main result with numerical simulations.
\end{abstract}


%%%%%%%%%%%%%%%%%%
\section{Introduction}
\IEEEPARstart{T}{he} benefits of employing advanced vehicular communication technologies and modern communication protocols have been expounded in great detail in recent literature; for example, Dedicated Short Range Communications (DSRC)~\cite{5888501}, Long Term Evolution (LTE)~\cite{8108463}, 5G~\cite{9345798}, Vehicle-to-Vehicle (V2V) communication~\cite{vinel2015vehicle}, etc. In addition to the use of onboard sensors (such as radars), advanced vehicular wireless communications can lead to a higher-level of connectivity that has the potential to significantly improve safety, increase mobility, and reduce fuel consumption \cite{darbha2018benefits}. In contrast to Adaptive Cruise Control (ACC), where the velocities and distances of adjacent vehicles are measured by onboard sensors, advanced wireless V2V communications can facilitate Cooperative Adaptive Cruise Control (CACC) where acceleration information of other vehicles can be employed to better predict the maneuvers of the neighboring vehicles and improve platooning safety and performance. 

For maintaining safe and tight formation to facilitate traffic throughput, the connected and autonomous vehicle platoon should exhibit string stability in addition to internal stability of each vehicle.  
To accomplish this, the key challenges that need to be addressed are: (i) selection of inter-vehicular spacing policy and control law synthesis, (ii) handling of imperfect information, and (iii) ensuring internal and string stability of the platoon in the presence of imperfect information. The spacing policy specifies the desired following distance;
%a function of the subject vehicle's speed; 
a constant time headway policy (CTHP) is typically employed in ACC and CACC systems, where the desired following distance is a linear function of the subject vehicle's speed with the proportional coefficient as the time headway~\cite{10.1115/1.2802497}. The communication mechanism, i.e., the information flow topology, provides the manner in which the motion information is exchanged in the platoon, and according to the direction of the information flow, communication can be unidirectional~\cite{RODONYI2019354}, bidirectional~\cite{KNORN2015208}, or ring-structured~\cite{10.1115/1.4036565}. It is common to employ unidirectional flow in vehicle platoons as it results in a simpler and yet robust communication and control structure.
 
By employing the formulated communication strategy, a control law is designed for robust string stability of the platoon to ensure robustness to imperfect information, uncertain dynamics, and external disturbances. 
%Concentrating on these problems, the research on autonomous and connected vehicle platoons has been attracting increasing efforts and numerous results have been reported, 
There has been extensive research on connected and autonomous vehicle platoons in recent decades in addition to the aforementioned papers, including e.g.~\cite{10026667,10074981,9939665,10041426,9989507,iet2020platooning,9457141,6683051,8573839,6515636,5571043,SILVA2021109542,1105955,9303465,9611165,9497781,7879221,GE2018445,BIAN201987,8673609,JIANG2021103110,9626600,9246221,9254117,9772756,SCHOLTE2022103511,doi:10.1287/trsc.2021.1100,9462542,8917522,GONG201825} and references therein. However, many existing results assume ideal V2V communication which is often unsuitable for practical implementation, as signal latency can be found in communicated information when using wireless communications. Those that do consider such imperfections do not provide bounds or control gains that can be readily employed for design and implementation of the algorithm. The discrepancy from the actual signal value from the ideal signal that is assumed for control design can not only degrade control performance but can also cause string instability due to the choice of time headway and control gains without consideration of signal delay in communicated signals. 

 This paper considers the following basic problem: Given the latency in V2V communication, what is a lower bound of the employable time headway for CACC and CACC+ based vehicle platooning as a function of the level of latency?   For this purpose, first, a CTHP decentralized control law with signal latency is considered and an inter-vehicular spacing error propagation equation for the platoon is derived which is utilized to derive a lower bound on the time headway for ensuring robust string stability in the presence of parasitic actuation lag and latency in communicated signals. Numerical simulation results on an example are provided to corroborate the main results. This paper provides an extension to our previous work in~\cite{darbha2018benefits} by considering signal latency in communicated information from the predecessor vehicle(s). Compared to the existing results, the main contributions of this paper are as follows:
\begin{itemize}
 \item Lower bounds on the time headway are derived for CACC and CACC+ based platooning that depend on the latency in the communicated information from the predecessor vehicle(s). 
 \item The analysis is conducted to facilitate systematic and step wise co-design procedure for the selection of the control gains as well as the time headway under a given latency and parasitic actuation lag. 
\end{itemize}

The remainder of the paper is organized as follows. Section~\ref{section:problem-formulation-and-preliminaries} contains the problem setup and preliminaries including vehicle dynamics and relevant definitions. The main results for CACC and CACC+ are provided in Section~\ref{section:main-results}. An illustrative numerical example and simulation results are provided in Section~\ref{section:numerical-simulations}. Concluding remarks and future work are given in Section~\ref{section:conclusion}.

\section{Preliminaries} \label{section:problem-formulation-and-preliminaries}
\begin{figure}[!htb]
\centering{\includegraphics[scale=0.6]{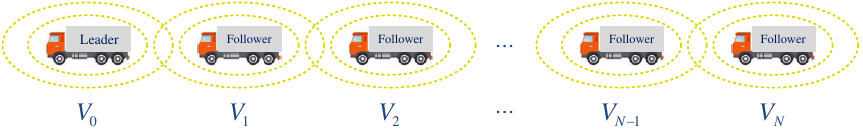}}
\caption{Connected and autonomous vehicle platoon with V2V communication.
\label{fig_1}}
\end{figure}

Consider a string of autonomous vehicles equipped with V2V communication as illustrated in Fig.~\ref{fig_1}, where $V_0$ denotes the leader vehicle and $V_1, V_2,  \cdots, V_{N - 1}, V_{N}$ denote the follower vehicles. %{\color{blue}(Add notation for $V_0$, $V_1$, $\cdots$)}
The dynamics of the $i$-th vehicle in the string are given by the model: 
\begin{align} \label{eq:vehicle-dynamics}
\left\{\begin{array}{*{20}{l}}
\ddot{x}_{i}(t) = a_{i}(t), \\
\tau \dot{a}_{i}(t) + a_{i}(t) = u_{i}(t),
\end{array}
\right.
\end{align}
where $x_{i}(t)$, $a_{i}(t)$, $u_{i}(t)$ represent the position, acceleration, and control input of the $i$-th vehicle at time instant $t$, and $i \in \mathcal{N} = \{ 1, 2, \cdots, N \}$, where $N$ is the total number of the following vehicles in the platoon, $\tau$ denotes the parasitic actuation lag. It is assumed that $\tau$ is {\it uncertain} with $\tau \in \left( 0, \tau_{0} \right]$, where $\tau_{0}$ is a positive real constant.
%
%\subsection{Definitions}
\begin{definition}[velocity dependent spacing error]
    Let $d$ denote the minimum or standstill spacing between adjacent vehicles, $v_i(t)$ denote the velocity of the $i$-th vehicle, $h_w$ denote the time headway, and $e_i(t) := x_{i}(t) - x_{i - 1}(t) + d$ be the spacing error for the $i$-th vehicle with respect to the $(i-1)$-th vehicle. Define the velocity dependent inter-vehicular spacing error for the $i$-th vehicle as: 
    \begin{align} \label{eq:delta-i-definition}
     \delta_i(t) = e_i(t) + h_w v_i(t).   
    \end{align}
\end{definition}

\begin{definition}[robust string stability~\cite{darbha2018benefits}] \label{definition:string-stability}
Let $H(s)$ denote the spacing error propagation transfer function that satisfies $\delta_{i}(s) = H(s) \delta_{i - 1}(s)$, $i \in \mathcal{N}$.
The connected and autonomous vehicle platoon is said to be robustly string stable if the following two conditions hold for every value of the parasitic lag, $\tau \in (0,\tau_0]$: (i) $H(s)$ achieves internal stability and (ii) the platoon achieves string stability, i.e., $\Vert \delta_{i}(t) \Vert_{\infty} \le \Vert \delta_{i - 1}(t) \Vert_{\infty}$, or in the frequency domain:
\begin{align} \label{eq:definition-2.1-H-j-omega-infinity-norm}
    \Vert H(j\omega) \Vert_{\infty} \le 1. %\forall \omega \ge 0. 
\end{align}
\end{definition}
For the CACC+ systems we consider here, Theorem 2 from \cite{9457141} would be helpful:
\begin{theorem}
    Suppose $$\delta_i(s) = \sum_{k=1}^r H_k(s) \delta_{i-k}(s), \forall i \ge 1, $$
\end{theorem}
with $\delta_k(s) \equiv 0$ if $k <0$ and $\delta_0(s)$ is the disturbance acting on the lead vehicle. Further, suppose that 
\begin{itemize}
    \item $\delta_0(s)$ corresponds to bounded and square-integrable lead vehicle's acceleration, 
    \item $$\sum_{k=1}^r \|H_k(j \omega)\|_{\infty} \le 1, $$
    \item There exists an $\alpha^* >0$ such that the $1-$norm of the initial error state of a vehicle platoon of any size is at most $\alpha^*$.
\end{itemize}
Then, there exists an $M_1, M_2 >0$ such that 
$$\|\delta_i(t)\|_{\infty} \le M_1 + M_2 \|\delta_0(t)\|_2.$$
A sufficient condition for 
$$\sum_{k=1}^r \|H_k(j \omega)\|_{\infty} \le 1$$ is that $\max_{1 \le k \le r} \|H_k(j \omega)\|_{\infty} \le \frac{1}{r}$, which we use here when analyzing CACC+ systems. 
\section{Main Results} \label{section:main-results}
In this section, we will consider a CTHP decentralized control law and derive the inter-vehicular spacing error propagation equation, and then proceed to derive the conditions on time headway under signal latency in the communicated information by considering internal and string stability. We will consider CACC first followed by CACC+.  

%%%%%%%%%%%%%%%%%%%%
\subsection{CACC with Communication Delay}
Considering delay in communicated acceleration with CACC, the control input for the $i$-th vehicle is given by 
\begin{align} \label{eq:control-input-CACC-with-delay-in-acceleration}
 u_i(t) = k_a a_{i - 1}(t - \ell) - k_v (v_i(t) - v_{i - 1}(t)) - k_{p} \delta_i(t),   
\end{align}
where $\ell$ is the communication delay and $k_a, k_v, k_p$ are positive gains. Then, under the above control law the inter-vehicular spacing error propagation equation is 
\begin{align} \label{eq:inter-vehicular-spacing-error-propagation-equation-1-CACC-with-latency}
 \delta_{i}(s) = H_1(s) \delta_{i - 1}(s),   
\end{align}
where $H_1(s) = \mathcal{N}_1(s)/\mathcal{D}_1(s)$ with $\mathcal{N}_1(s) = k_a s^2 e^{- \ell s} + k_v s + k_{p}$ and $\mathcal{D}_1(s) = \tau s^3 + s^2 + (k_v + h_w k_{p}) s + k_{p}$. By way of notation, we will sometimes use $H_1(s; \tau)$ to emphasize the dependence of the spacing error propagation transfer function, $H_1$, on the parameter $\tau$. The following theorem provides a lower bound on the minimum time headway in the presence of communication delay. 
\begin{theorem} \label{theorem:string-stability-condition-CACC-with-latency}
The following are true for the platoon given by the inter-vehicular spacing error propagation equation~\eqref{eq:inter-vehicular-spacing-error-propagation-equation-1-CACC-with-latency}:
\begin{enumerate}[(a)] 
\item $\| H_1(j \omega; \tau) \|_{\infty} \le 1$, $\forall \tau \in (0, \tau_0]$, implies that $k_a \in (0,1)$; 
\item given any $k_a \in (0,1)$, $\eta > 0$, and a value of $h_w$ satisfying  
\begin{align} \label{eq:hw-lower-bound-CACC-with-latency}
  h_w = \max\left\{\frac{2 (\tau_0 + k_a \ell) }{1 + k_a}, \frac{\ell}{2}\right\}(1+\eta),
 \end{align}
there exist $k_p, k_v > 0$ such that  $\| H_1(j \omega) \|_{\infty} \le 1$ for all $\tau \in (0,\tau_0]$.
\end{enumerate}
\end{theorem}
\begin{IEEEproof}
Let $\gamma := k_v + h_w k_p$. For internal stability, the characteristic equation $\mathcal{D}_1(s)=0$ is Hurwitz $\forall \tau \in (0, \tau_0]$ iff  
\begin{align} \label{eq:int-stab-cond}
 \gamma - \tau_0 k_{p} > 0.   
\end{align}
\noindent (a) Consider $H_1(j \omega) = \mathcal{N}_1(j \omega)/\mathcal{D}_1(j \omega)$ where 
 \begin{align}
 \mathcal{N}_1(j \omega) &= - k_a \omega^2 e^{-j \omega \ell} + j \omega k_v + k_{p} \nonumber \\
 &= j \omega ( k_a \omega \sin(\omega \ell) + k_v) + k_{p} - k_a \omega^2 \cos(\omega \ell), \nonumber \\
 \mathcal{D}_1(j \omega) &= j \omega (\gamma - \tau \omega^2) + k_{p} - \omega^2.
 \end{align}  
 We will show the contrapositive of statement (a): if $k_a \ge 1$, then $\| H_1(j \omega; \tau) \|_{\infty} > 1$ for some $\tau \in (0, \tau_0]$ through the following steps. \\
 Case (1) $k_a > 1$: We will only deal with the case $\ell \ne 0$ as $\ell=0$ is considered in~\cite{konduri2017robust}. We will show that there exists a $\hat{\omega}$ such that (1) $\hat{\omega} \ell = 2 k \pi$ for some sufficiently large positive integer $k$, (2)  $\tau\hat{\omega}^2 = \gamma$ for some $\tau \in (0,\tau_0]$, and (3) $\| H_1(j \hat{\omega}; \tau) \|_{\infty} > 1$ for some $\tau \in (0, \tau_0]$. We will select $\hat{\omega}$ as follows to satisfy conditions (1) and (2): pick an integer $k$ large enough such that $k^2 > \frac{\gamma \ell^2}{4 \pi^2 \tau_0}$; define $\hat{\omega} := \frac{2 k \pi}{\ell}$ so that $\hat{\omega} \in \left[ \sqrt{ \frac{\gamma}{\tau_0} }, \infty \right)$; pick $\tau$ such that $\tau\hat{\omega}^2 = \gamma$. Note that $\tau \in (0, \tau_0]$. Now we will focus on showing condition (3). 
 %In this case, one can find a natural number $k$ satisfying $k^2 \ge \frac{\gamma \ell^2}{4 \pi^2 \tau_0}$. Let $\hat{\omega} := \sqrt{\frac{\gamma}{\tau}} \in \left[ \sqrt{ \frac{\gamma}{\tau_0} }, \infty \right)$ so that $\hat{\omega} \ell = 2 k \pi$.  
%

If $k_a > 1$ and the internal stability condition~\eqref{eq:int-stab-cond} is satisfied, i.e., $\gamma - \tau_0 k_p > 0$ implies $\gamma - \tau k_p > 0$, then we have $\hat{\omega}^2 > k_{p} > \frac{k_{p}}{k_a}$. Note that   
\begin{align} \label{eq:H1-norm-inequality-1}
 \| H_1(j \omega; \tau) \|_{\infty} &\ge \left\lvert \frac{ j \hat{\omega} ( k_a \hat{\omega} \sin(\hat{\omega} \ell) + k_v ) + k_{p} - k_a \hat{\omega}^2 \cos(\hat{\omega} \ell) }{ k_{p} - \hat{\omega}^2 }\right\rvert \nonumber
 \end{align}
Substituting $\hat{\omega}\ell = 2 k \pi$, we get 
 \begin{align}
 \| H_1(j \omega; \tau) \|_{\infty} & \ge \left\lvert \frac{j\hat{\omega} k_v + k_{p} - k_a \hat{\omega}^2}{k_{p} - \hat{\omega}^2} \right\rvert \nonumber \\
 &> k_a \left\lvert \frac{ \frac{k_{p}}{k_a} - \hat{\omega}^2 }{k_{p} - \hat{\omega}^2} \right\rvert > k_a > 1.  
\end{align}
Case (2) $k_a = 1$: In this case, %if $k_a = 1$, 
it suffices to show that there exists an $\omega$ and a $\tau \in (0, \tau_0]$ for which  $\| H_1(j \omega; \tau) \|_{\infty} > 1$. Note that
 \begin{align}\label{eq:case2-ka1}
 \| H_1(j \omega; \tau) \|_{\infty}^2 \ge \frac{ \omega^2 ( \omega \sin(\omega \ell) + k_v )^2 + ( k_{p} - \omega^2 \cos(\omega \ell) )^2 }{ \omega^2 ( \gamma - \tau \omega^2)^2 + (k_{p} - \omega^2)^2 }.    
 \end{align}

 As before, if we were to restrict $\omega = \frac{2k\pi}{\ell}$ in the right hand side of the inequality \eqref{eq:case2-ka1}, we see that it suffices to find an integer $k$ and a $\tau \in (0, \tau_0]$ such that 
 $$ \frac{\omega^2k_v^2 + (k_p - \omega^2)^2}{\omega^2(\gamma - \tau \omega^2)^2+(k_p - \omega^2)^2} >1,$$ 
 or equivalently, it suffices to find a $\omega$ such that $\omega \ell = 2k \pi$ for some integer $k$ and a $\tau \in (0, \tau_0]$ satisfying
 $$ k_v^2 > (\gamma - \tau \omega^2)^2.$$
 Since we are restricting $\omega \ell = 2k \pi$, what we are seeking may be recast as follows: Find an integer $k$ and a $\tau \in (0, \tau_0]$ satisfying 
 $$ -k_v < \gamma - \tau \omega^2 < k_v \iff -k_v < k_v + h_w k_p - \tau \frac{4k^2\pi^2}{\ell^2} < k_v $$
This is equivalent to showing that there exists an integer $k$ and a $\tau \in (0,\tau_0]$ satisfying
\begin{align*}
k_ph_w &< \tau k^2\frac{4\pi^2}{\ell^2} < k_ph_w + 2k_v \\
& \iff k_ph_w \frac{\ell^2}{4\pi^2}< \tau k^2 < (k_ph_w + 2k_v)\frac{\ell^2}{4\pi^2} 
\end{align*}
 
%  Define $\tilde{\omega}:= \frac{2k \pi}{\ell} 
%  $.
% Then 
% \begin{align*}
%     h_w k_p &< \tau \tilde{\omega}^2 < h_w k_p + 2 k_v, \\
%  0 &< \tau \tilde{\omega}^2 - h_w k_p  < 2 k_v, \\
%  -2k_v &< h_w k_p - \tau \tilde{\omega}^2  < 0, \\ 
%  -k_v &< \gamma - \tau \tilde{\omega}^2  < k_v.
% \end{align*}
% This implies that $k_v^2 > (\gamma - \tau \tilde{\omega}^2)^2$. Since $\sin(\tilde{\omega} \ell) = 0, \; \; \cos(\tilde{\omega} \ell) = 1$, and $k_v^2 > (\gamma - \tau \tilde{\omega}^2)^2$,  we then have
% \begin{align} \label{eq:H1-norm-square-ka-equal-to-1}
%  \| H_1(j \omega; \tau) \|_{\infty}^2 &\ge \frac{ \tilde{\omega}^2 k_v^2 + ( k_{p} - \tilde{\omega}^2 )^2 }{ \tilde{\omega}^2 ( \gamma - \tau \tilde{\omega}^2 )^2 + ( k_{p} - \tilde{\omega}^2 )^2 } \nonumber \\
%  &> 1.
% \end{align}
%
We will first check if the interval 
$$(\frac{\ell^2}{4\pi^2\tau_0} k_ph_w, \frac{\ell^2}{4\pi^2 \tau_0} (2k_v+ k_ph_w))$$
contains the square of an integer, say $k^2$; if this is true, then we set $\tau = \tau_0$. If not, we can scale the above interval by choosing a $\tau \in (0, \tau_0]$ so that the interval 
$$\left(\frac{\ell^2}{4\pi^2\tau} k_ph_w, \frac{\ell^2}{4\pi^2 \tau} (2k_v+ k_ph_w)\right)$$
contains the square of an integer, say $k^2$. 
In either case, the desired integer $k$ and $\tau$ satisfying the condition $k_v^2 > (\gamma - \tau \omega^2)^2$, and hence, $\| H_1(j \omega; \tau) \|_{\infty} > 1$. \\

\noindent (b) First, we observe that  
$$\| H_1(j \omega; \tau) \|_{\infty}^2 = \sup_{\omega \ge 0} (\vert \mathcal{N}_1(j \omega) \vert^2/\vert \mathcal{D}_1(j \omega) \vert^2)$$ 
where 
\begin{align*}
\vert \mathcal{N}_1(j \omega) \vert^2 &= k_a^2 \omega^4 + 2 k_a k_v \omega^3 \sin (\omega \ell) \\
&\qquad + (k_v^2 - 2 k_a k_{p} \cos(\omega \ell) ) \omega^2 + k_{p}^2, \\
\vert \mathcal{D}_1(j \omega) \vert^2 &= \tau^2 \omega^6 + (1 - 2 \tau \gamma) \omega^4 \\
&\qquad + (\gamma^2 - 2 k_{p}) \omega^2 + k_{p}^2. 
\end{align*}

The condition $\| H_1(j \omega; \tau) \|_{\infty}^2 \le 1$ is equivalent to 
$$\vert \mathcal{D}_1(j \omega) \vert^2 - \vert \mathcal{N}_1(j \omega) \vert^2 \ge 0, \quad \forall \omega \ge 0,$$ 
that is,  
 \begin{align} \label{eq:string-stability-equivalent-inequality-1}
 & \tau^2 \omega^4 + (1 - k_a^2- 2 \tau \gamma) \omega^2 + \gamma^2 - 2 k_{p} - k_v^2 \nonumber \\
 & \quad - 2 k_a k_v \omega \sin(\omega \ell) +  2 k_a k_{p} \cos(\omega \ell) ) \ge 0.
 \end{align}
The following inequalities always hold for the sine and cosine functions:  
\begin{align}\label{eq:sine-cosine-bounds}
  \begin{cases}{}
 - \omega \ell \le \sin(\omega \ell) \le \omega \ell, \\
 1 - \frac{\omega^2 \ell^2}{2} \le \cos(\omega \ell) \le 1.
  \end{cases}
 \end{align}
Then, substituting \eqref{eq:sine-cosine-bounds},  inequality~\eqref{eq:string-stability-equivalent-inequality-1} is satisfied if  
\begin{align}\label{eq:string-stability-equivalent-inequality-2}
\tau^2 \omega^4 &+ (1 - 2 \tau \gamma - k_a^2 - 2 k_a k_v \ell - k_a k_p \ell^2) \omega^2 \nonumber \\
&+ \gamma^2 - ( 2 k_{p} - 2 k_a k_{p} + k_v^2 ) \ge 0. 
\end{align}
Inequality \eqref{eq:string-stability-equivalent-inequality-2} is satisfied if 
\begin{subnumcases}{}
 1 - 2 \tau \gamma - k_a^2 - 2 k_a k_v \ell - k_a k_p \ell^2  \ge 0, \label{eq:string-stability-equivalent-inequality-condition-a} \\
 \gamma^2 - ( 2 k_{p} - 2 k_a k_{p} + k_v^2 ) \ge 0. \label{eq:string-stability-equivalent-inequality-condition-b}
\end{subnumcases}
Using $k_v = \gamma - h_w k_p$ in~\eqref{eq:string-stability-equivalent-inequality-condition-a} and rearranging terms,  ~\eqref{eq:string-stability-equivalent-inequality-condition-a} can be rewritten as   
 \begin{align} \label{eq:gamma-upper-bound-inequality}
1 - k_a^2 - (2 \tau + 2 k_a \ell) \gamma + 2 k_a k_p \ell \left(h_w - \frac{\ell}{2} \right) \ge 0.  
 \end{align}
Since $k_a\in (0,1)$ and $h_w > \frac{\ell}{2}$, \eqref{eq:gamma-upper-bound-inequality} is satisfied if 
 \begin{align}
 \gamma \le \frac{1 - k_a^2}{2 \tau_0 + 2 k_a \ell}. \label{eq:gamma-upper-bound-2}
 \end{align} 
In addition, from~\eqref{eq:string-stability-equivalent-inequality-condition-b}, we have  
 \begin{align} \label{eq:gamma-lower-bound-1}
 \gamma \ge \sqrt{ 2 k_{p} (1 - k_a) + k_v^2 }.  
 \end{align}
 Combining~\eqref{eq:gamma-upper-bound-2} and~\eqref{eq:gamma-lower-bound-1} yields 
\begin{align} \label{eq:gamma-combined-inequality-cacc} 
 \sqrt{ 2 k_{p} (1 - k_a) + k_v^2 } \le \gamma \le \frac{1 - k_a^2}{2 \tau_0 + 2 k_a \ell}.
\end{align}
For $\|H_1(j\omega; \tau\| \le 1$ for all $\tau \in (0, \tau_0]$, we require inequality \eqref{eq:gamma-combined-inequality-cacc} to hold. 
Next, we will show the existence of $k_v$ and $k_{p}$ satisfying \eqref{eq:gamma-combined-inequality-cacc}. 
Substituting $\gamma = k_v+h_wk_p$ in $\gamma^2 \ge 2 k_{p} (1 - k_a) + k_v^2$ and simplifying, we obtain 
\begin{align}
    2 h_w k_v + h_w^2 k_p \ge 2 (1-k_a),
\end{align}
which can be rewritten as
 \begin{align}\label{eq:set-S1-cond}
 \frac{k_v}{a_1} + \frac{k_{p}}{b_1} \ge 1, 
 \end{align} 
where 
 \begin{align}
 a_1 &= 
 \frac{1 - k_a}{h_w}, \ 
 b_1 = \frac{2(1 - k_a)}{h_w^2}.
 \end{align}
 Then, the admissible values of $k_v$ and $k_{p}$ satisfying \eqref{eq:set-S1-cond} are given by:
 \begin{align}
 \mathcal{S}_1 \coloneqq \left\{ (k_v, k_{p}): k_v > 0, k_{p} > 0, \frac{k_v}{a_1} + \frac{k_{p}}{b_1} \ge 1  \right\}.
 \end{align}
Similarly, from $\gamma \le \frac{1 - k_a^2}{2 (\tau_0 + k_a \ell)}$, we have
 \begin{align}
 k_v + h_w k_{p} \le \frac{1 - k_a^2}{2 (\tau_0 + k_a \ell) },
 \end{align}
which can be rewritten as 
\begin{align}\label{eq:set-S2-cond}
 \frac{k_v}{a_2} + \frac{k_{p}}{b_2} \le 1,
 \end{align}
 where 
 \begin{align}
 a_2 = \frac{1 - k_a^2}{2 (\tau_0 + k_a \ell)}, \ b_2 = \frac{1 - k_a^2}{2(\tau_0 + k_a \ell) h_w}.
 \end{align}
Then, the admissible values of $k_v$ and $k_{p}$ satisfying \eqref{eq:set-S2-cond} are given by: 
 \begin{align}
 \mathcal{S}_2 \coloneqq \left\{ (k_v, k_{p}): k_v > 0, k_{p} > 0, \frac{k_v}{a_2} + \frac{k_{p}}{b_2} \le 1 \right\}. 
 \end{align}
Then, all the admissible values of $k_v$ and $k_{p}$ satisfying \eqref{eq:gamma-combined-inequality-cacc} are given by $\mathcal{S} = \mathcal{S}_1 \cap \mathcal{S}_2$. 
For $\mathcal{S}$ to be non-empty, the following equation  
\begin{align} \label{eq:kvkp_equation}
    \underbrace{\begin{bmatrix}
        \frac{1}{a_1} & \frac{1}{b_1} \\ \frac{1}{a_2} & \frac{1}{b_2}
    \end{bmatrix}}_{A}
    \begin{bmatrix}
        k_v \\ k_p
    \end{bmatrix}
    = \begin{bmatrix}
        1 \\ 1
    \end{bmatrix}
\end{align}
should have a solution in quadrant I or II in the $k_v$ (horizontal axis) - $k_p$ (vertical axis) plane, i.e. $k_p > 0$ in~\eqref{eq:kvkp_equation}.
Note that $A$ is invertible because $\det A  = \frac{(1-k_a)(1-k_a^2)}{2(\tau_0+k_a \ell)h_w^2} > 0$. We can obtain the soultion for~\eqref{eq:kvkp_equation} as follows: 
\begin{align}
    \begin{bmatrix}
     k_v \\ 
     k_p
    \end{bmatrix} = 
     \begin{bmatrix}
     \frac{ a_1 a_2 ( b_1 - b_2 ) }{ a_2 b_1 - a_1 b_2 } \\
      \frac{  b_1 b_2 ( a_2 - a_1 )}{ a_2 b_1 - a_1 b_2 }
    \end{bmatrix}.
\end{align}
For $k_p>0$, we require $a_1 < a_2$, i.e., 
\[
 \frac{ 1 - k_a }{ h_w } < \frac{ 1 - k_a^2 }{ 2 ( \tau_0 + k_a \ell ) },
\]
from which we obtain:
\begin{align} \label{eq:implemented-hw-CACC}
 h_w > \frac{ 2 (\tau_0 + k_a \ell) }{1 + k_a}.
 \end{align}
%
%Let $h_w$ be as given in \eqref{eq:hw-lower-bound-CACC-with-latency}, 
%which satisfies \eqref{eq:implemented-hw-CACC}, then $b_1 > b_2$, which implies $k_v > 0$. 

%%%
Since the condition for the set $\mathcal{S}$ to be non-empty is satisfied by the chosen time headway according the hypothesis of the theorem, there exist control gains $k_p, k_v > 0$ such that  $\| H_1(j \omega) \|_{\infty} \le 1$ for all $\tau \in (0,\tau_0]$.

Finally, the following shows that the internal stability condition is satisfied when $h_w$ satisfies~\eqref{eq:hw-lower-bound-CACC-with-latency}.
    \begin{align}
        \gamma &= k_v + h_w k_p \nonumber\\
        & \ge k_v + \frac{ 2 (\tau_0 + k_a \ell) }{1 + k_a} ( 1 +\eta ) k_p \nonumber \\
        & > k_v + \frac{ 2 ( \tau_0 + k_a  \ell ) }{ 1 + k_a } k_p \nonumber \\
        & = \frac{2\tau_0 k_p}{1+k_a} + \frac{2 k_a k_p \ell}{1+k_a}  + k_v \nonumber \\
        & > \frac{2\tau_0 k_p}{1+k_a}.
    \end{align}
Since $1+k_a < 2$, we have $\gamma > \tau_0 k_{p}$.
\end{IEEEproof}

The following remarks provide insights into the proposed CACC control design with communication delay and also highlight the points of departure from the existing work. 
\begin{remark}
    {\em Comparison of CACC with ACC}: If we consider 
    $$\bar{\tau}(k_a) = \max\left\{\frac{1}{1+k_a} \tau_0 + \frac{k_a}{1+k_a} {\ell}, \frac{\ell}{4} \right\}$$
    we can think of effective lag as the convex combination of  the maximum parasitic lag, $\tau_0$ and communication delay, $\ell$. The bound becomes $h_w \ge 2{\bar \tau} = h_{w, \min}$. The weights depend on the choice of $k_a$; clearly, when $k_a=0$, (corresponding to ACC), we recover the result of \cite{swaroop_PhD_thesis}. We also note that $\bar \tau \ge \min\{\ell, \tau_0\}$, and it is better to bias $k_a$ appropriately to lower the bound for $h_{w}$.
\end{remark}
%%%%%%%%%%%%%%%%%%%%
\subsection{CACC+ with Communication Delay}
For CACC+ with information from $r$ predecessor vehicles, the following control input to the $i$-th vehicle is considered: 
\begin{align} \label{eq:control-law-r-predecessors}
 u_i(t) &= k_{a} a_{i - 1}(t - \ell) - k_{v} (v_i(t) - v_{i - 1}(
t)) - k_{p} \delta_i(t) \nonumber \\
& \quad + \sum\limits_{q = 2}^{r} \left[ k_{a} a_{i - q}(t - \ell) - k_{v} \left(v_i(t) - v_{i - q}(t - \ell) \right) \right. \nonumber \\  
 &\qquad\qquad \left. - k_{p} \left(x_i(t) - x_{i - q}(t - \ell) + d_q + q h_{w} v_i(t) \right) \right],
\end{align}
where $\ell$ denotes the delay, $d_q = q d$, and $k_{a}, k_{p}, k_{v}$ are the controller gains. The resulting inter-vehicular spacing error propagation equation for the $i$-th vehicle is given by
\begin{align} \label{eq:intervehicular-spacing-error-propagation-equation-r-predecessors}
 \delta_i(s) = \sum\limits_{q = 1}^{r} H_{q}(s) \delta_{i-q}(s), \ i > r
\end{align}
where $H_{q}(s) = \mathcal{N}_{q}(s)/\mathcal{D}_r(s)$, in which
\begin{align*}
    \mathcal{N}_{1}(s) &= k_{a} s^2 e^{- \ell s} + k_{v} s + k_{p}, \\
    \mathcal{N}_{q}(s) &= e^{- \ell s} (k_{a} s^2 + k_{v} s + k_{p}), \ q = 2, \cdots, r, \\
    \mathcal{D}_r(s) &= \tau s^3 + s^2 + \sum\limits_{q = 1}^{r} \left[ ( k_{v} + q h_{w} k_{p} ) s + k_{p} \right].
\end{align*} 
Let $\bar{k}_v := rk_v, \bar{k}_a := rk_a, \bar{k}_p := r k_p$, $\bar{h}_w=\frac{r+1}{2} h_w$, and  $\gamma_r := r k_{v} + \frac{r(r + 1)}{2} h_{w} k_{p} = \bar{k}_v + \bar{h}_w \bar{k}_p$. Then, $\mathcal{D}_r(s)$ can be written as 
\[
\mathcal{D}_r(s) = \tau s^3 + s^2 +  \gamma_r s + \bar{k}_{p}.
\]
The following theorem quantifies the main result for the CACC+ case.
\begin{theorem} \label{theorem:CACC+identical-gain-case}
For the CACC+ platoon given by the inter-vehicular spacing error propagation equation~\eqref{eq:intervehicular-spacing-error-propagation-equation-r-predecessors}, the following are true:
\begin{enumerate}[(a)]
    \item $\max_{1\leq q \leq r} \vert H_{q}(j \omega) \vert \le \frac{1}{r}$, $\forall \omega \ge 0$, implies that $\bar{k}_{a} \in (0,1)$; 
    \item given any $\bar{k}_{a} \in (0,1)$, $\eta_r > 0$ and $h_{w}$ satisfying 
        \begin{align}
            \bar{h}_{w} = \frac{2 (\tau_0 + \bar{k}_a \ell) }{1 + \bar{k}_a}(1+\eta_r),   
        \end{align}
    there exist $\bar{k}_{v} > 0$ and $\bar{k}_{p} > 0$ such that, for all $\tau \in (0, \tau_0]$,   $r \vert H_{q}(j \omega) \vert \le 1$.  
  \end{enumerate}
\end{theorem}
\begin{IEEEproof}
Note that 
\begin{eqnarray*}
    \bar{H}_1(s; \ell) :=rH_1(s) = \frac{\bar{k}_a e^{- \ell s} + {\bar k}_v s + {\bar k}_p}{\tau s^3 + s^2 + (\bar{k}_v + {\bar h}_w \bar{k}_p)s + {\bar k}_p}.
\end{eqnarray*}
Note that $\bar{H}_1(s;\ell)$ is identical in structure to  $H_1(s)$ considered in Theorem 3.1. Applying Theorem 3.1 to  $\bar{H}_1(s)$, we obtain that 
$\|\bar{H}_1(j \omega) \| \le 1, \; \forall \tau \in (0, \tau_0]$ implies that ${\bar k}_a \in (0,1)$; furthermore, given any ${\bar k}_a \in (0,1), \; \eta_r >0,$ and a value of ${\bar h}_w$ satisfying 
${\bar h}_w^{(1)} = \frac{2(\tau_0 + \bar{k}_a \ell)}{1+ \bar{k}_a}(1+ \eta_r).$

Moreover,  for $q \ge 2$, 
$\|{\bar H}_1(s; 0)\| = \|r H_q(j \omega)\|$. One may still apply Theorem 3.1 to the transfer function $r H_q(s), \; q \ge 2$. However, the corresponding lower bound will be 
${\bar h}_w^{(2)} = \frac{2\tau_0}{1+ \bar{k}_a}(1+ \eta_r).$

Since both bounds must be satisfied, picking ${\bar h}_w \ge \max \{{\bar h}_w^{(1)}, {\bar h}_w^{(2)}\} = {\bar h}_w^{(1)}$ suffices to guarantee string stability. 

% The proof of this theorem is provided in {\sc Appendix~\ref{Appendix:Proof-Theorem-3.2}} and follows along the same lines as the proof of the CACC case.
\end{IEEEproof}

\begin{remark}
    The condition given in terms of $\bar{h}_w$ by equation (34) can be restated in terms of the time headway ($h_w$) as:
    $$h_w = \frac{4(\tau_0 + rk_a \ell)}{(r+1)(1+rk_a)}(1+ \eta_r).$$
    Even with a fixed communication lag (imperfect communication), the mobility and safety benefits accruing from the information of $r$ immediately preceding vehicles are clear through a reduction in $h_w$.
\end{remark}

\begin{remark}
   This work differs from the work of \cite{9462542} in the following ways:
   \begin{itemize}
    \item    In~\cite{9462542}, the time headway selection problem for connected and autonomous vehicles with communication delay was considered;
   moreover, the feasibility of system of inequalities in Theorem 2 (see inequalities (29a)-(29h)) crucial for their main result was not addressed.
       \item Vehicle model in this paper considers $\tau$ to be an uncertain parameter which can take any value in the interval $(0, \tau_0]$. Robustness results presented here allow for this variation and hold for every $\tau \in (0, \tau_0]$. In contrast, reference \cite{9462542} requires $\tau, \ell$ to be related as $rk_a \ell \le \tau$; clearly, for given $r,  k_a, \ell$,  the above inequality cannot hold when $\tau$ is sufficiently small. This relates to the above remark on the feasibility of inequalities (29a)-(29h) of their Theorem 2. 
       \item The control law in \cite{9462542} uses acceleration feedback of the ego vehicle in contrast to the work presented here, which does not use acceleration feedback. 
       \item The results presented here recover earlier results; setting $\ell=0$, we can recover the results of \cite{konduri2017robust}. When $r=1$ and $k_a = 0$, we recover the result of \cite{swaroop_PhD_thesis} for ACC systems. 
   \end{itemize}
\end{remark}
\begin{remark}
The approach presented in this paper allows for an implementation that bootstraps CACC+ on top of CACC which in turn can be bootstrapped on ACC. It will be easier to transition from CACC+ to CACC to ACC in the event of communication disruptions or failures, and thus, provides robustness and a degree of safety.
\end{remark}
\section{Numerical Simulations} \label{section:numerical-simulations} 
 In this section, we present numerical simulations on an example to illustrate the main results. We consider the following numerical values for the system parameters: $\tau_0 = 0.5$ s, $d = 5$ m, $N = 12$, and the steady-state velocity of the vehicles to be 25 m/s. In the numerical simulation, the worst case $\tau$ value was chosen, i.e., $\tau = \tau_0$. We assume that the lead vehicle experiences an external disturbance which causes a perturbation in its acceleration, denoted by $a_0(t)$, given by: 
\begin{align}
 a_0(t) = 
 \begin{cases}
  0.5 \sin (0.1 (t - 10)), \ 10 < t < 20\pi + 10, \\
  0, \ \mbox{otherwise}.
 \end{cases}   
\end{align}
For this scenario, we evaluate the performance under the two communication and control strategies considered in Section~\ref{section:main-results} in the following subsections.
%%%%%%%%%%%%%%%%%%%%
%
\subsection{CACC with Communication Delay}
\begin{figure}[!htb]
\centering{\includegraphics[scale=0.5]{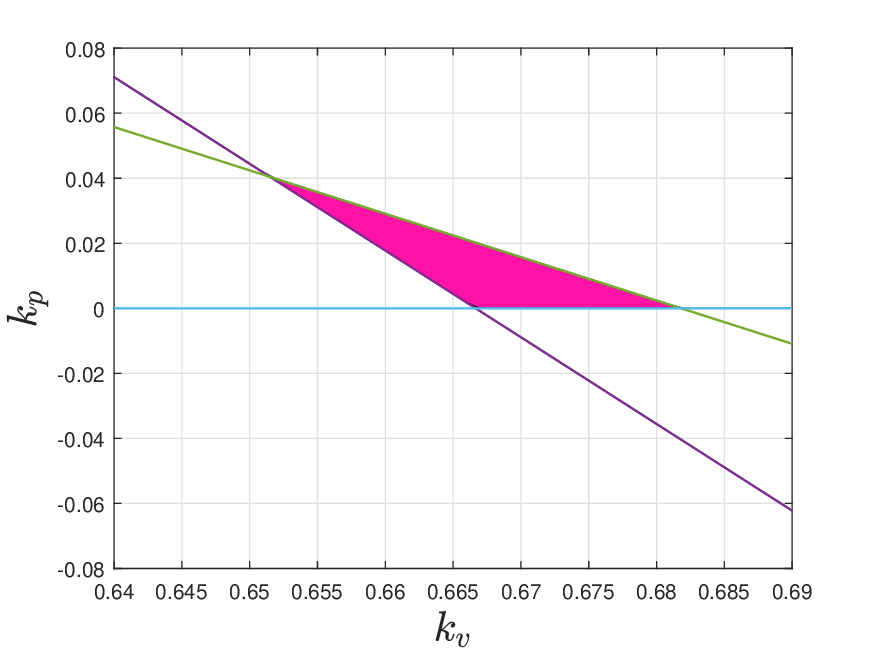}}
\caption{($k_v, k_p$) feasible region for the CACC case ($k_a = 0.5$, $h_w = 0.75$ s).
\label{fig:k_v_k_p_region_CACC}}
\end{figure}

Let $k_a = 0.5$ and assume $\ell = 100$ ms. From~\eqref{eq:implemented-hw-CACC}, we have 
 \begin{align}
 h_w > \frac{2 (\tau_0 + k_a \ell) }{1 + k_a} = 0.7333 \ \mbox{s}.     
 \end{align}
Choosing $h_w = 0.75$ s, the parameters $a_1, b_1, a_2$ and $b_2$ are given by
 \begin{align}
  a_1 = 0.6667, b_1 = 1.7778, a_2 = 0.6818, b_2 = 0.9091.   
 \end{align}
 The feasible region of $(k_v, k_p)$ determined by $\frac{k_v}{a_1} + \frac{k_p}{b_1} \ge 1$ and $\frac{k_v}{a_2} + \frac{k_p}{b_2} \le 1$ is shown in Fig.~\ref{fig:k_v_k_p_region_CACC}.  
Choosing $k_v = 0.67$, we can obtain $0 < k_p \le 0.0158$, from which we choose $k_p = 0.014$.

\begin{figure}[!htb]
\centering{\includegraphics[scale=0.5]{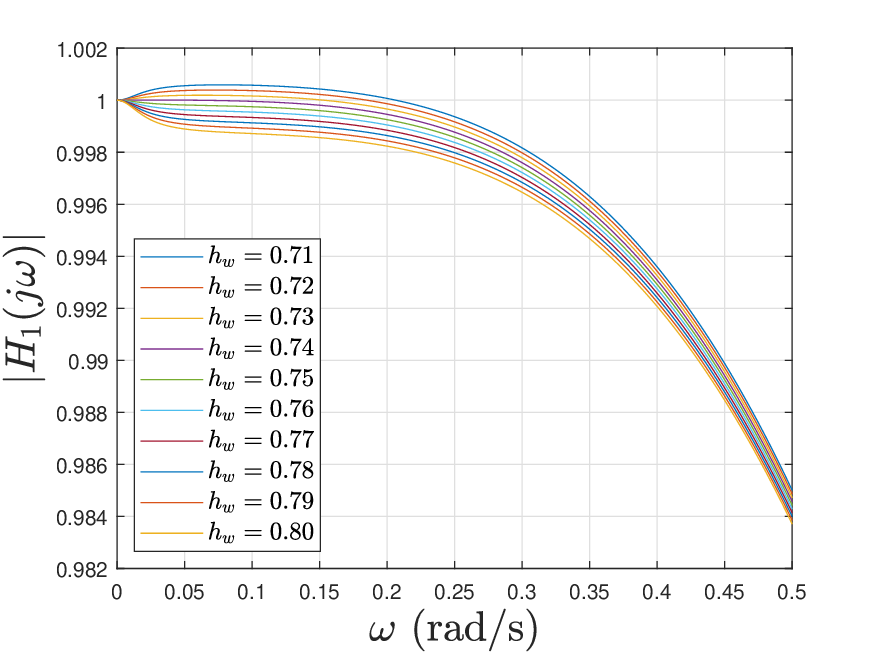}}
\caption{Comparison of $\vert H_1(j \omega) \vert$ for different $h_w$.
\label{fig:PF-latency-hw-pecision}}
\end{figure}

\begin{figure}[!htb]
\centering{\includegraphics[scale=0.5]{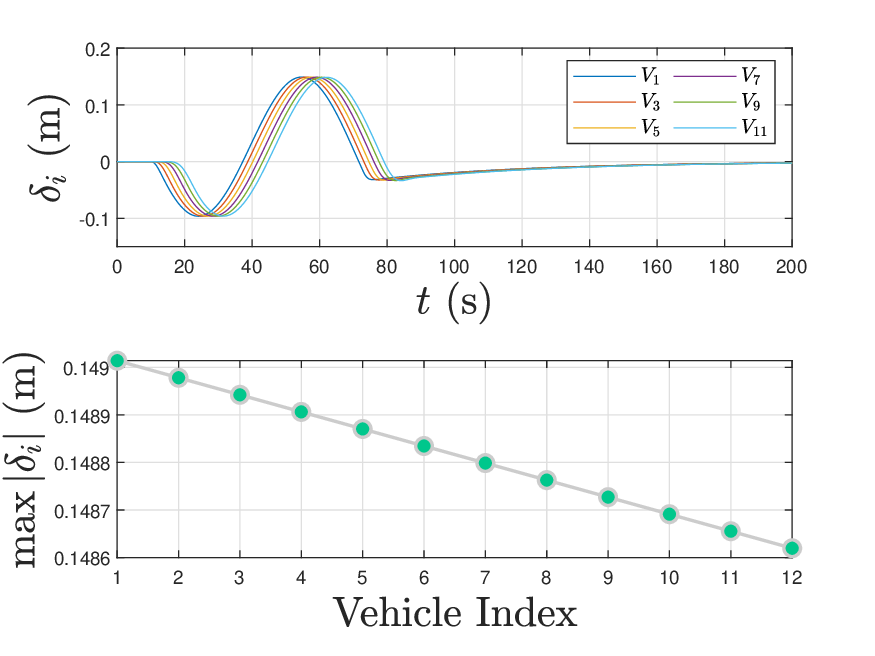}}
\caption{Inter-vehicular spacing errors (CACC, $h_w = 0.75$ s).
\label{fig:PF-latency-delta}}
\end{figure}

\begin{figure}[!htb]
\centering{\includegraphics[scale=0.5]{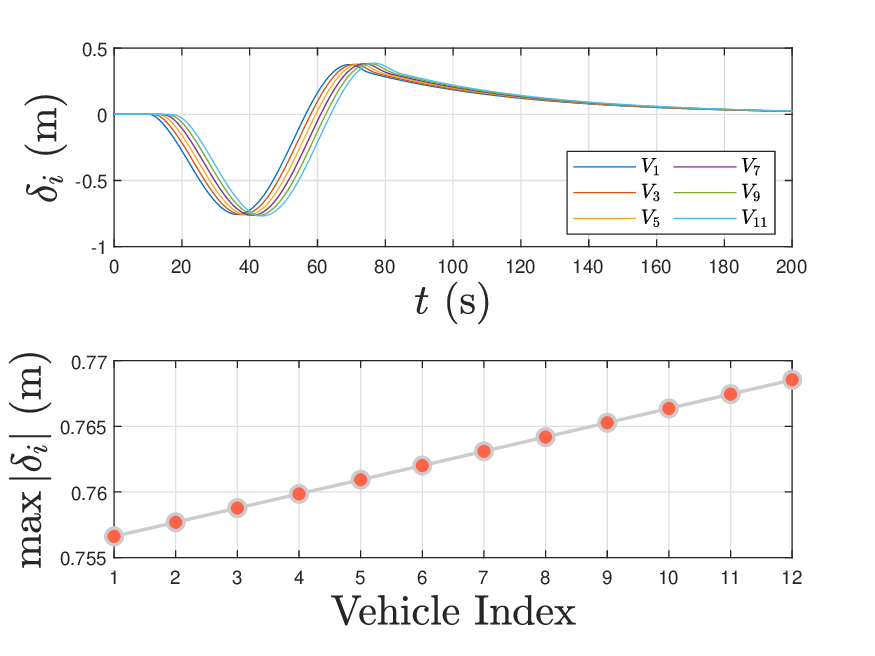}}
\caption{Inter-vehicular spacing errors (CACC, $h_w = 0.65$ s).
\label{fig:CACC_comparison}}
\end{figure}

The comparison of $\vert H_1(j \omega) \vert$ for  different values of $h_w$ is provided in Fig.~\ref{fig:PF-latency-hw-pecision}. 
With the above determined values of the control gains and time headway, the evolution of the inter-vehicular spacing errors is shown in Fig.~\ref{fig:PF-latency-delta}.  
In contrast, when $h_w = 0.65$ s, which is less than the lower bound 0.7333 s, the evolution of the inter-vehicular spacing errors for this $h_w$ value is provided in Fig.~\ref{fig:CACC_comparison}, which exhibits string instability.
%%%%%%%%%%%%%%%%%%%%
\subsection{CACC+ with Communication Delay}
\begin{figure}[!htb]
\centering{\includegraphics[scale=0.5]{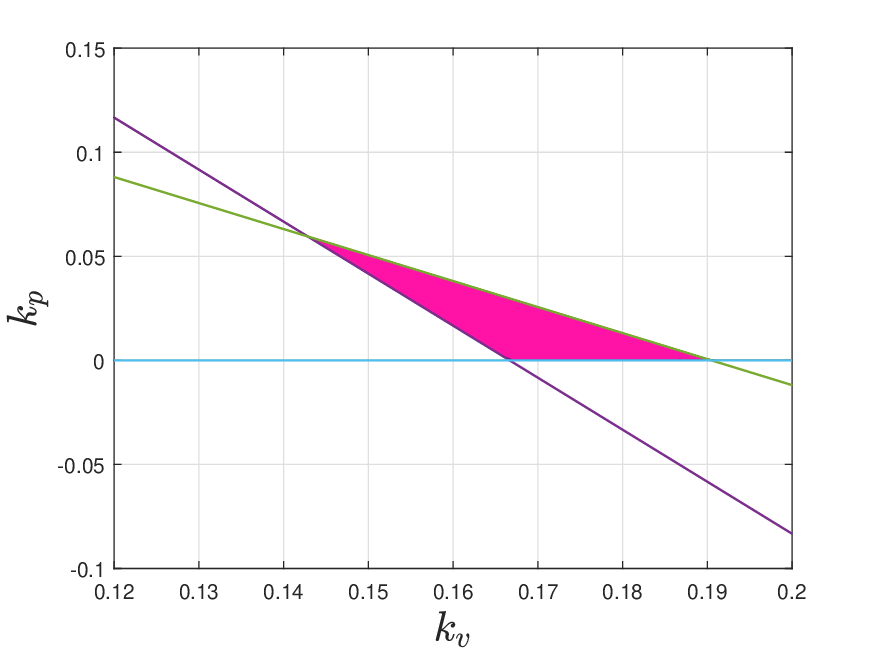}}
\caption{($k_v, k_p$) feasible region for the CACC+ case ($k_a = 0.2$, $h_w = 0.4$ s).
\label{fig:k_v_k_p_region_CACC_plus}}
\end{figure}

Let $r = 3$ and assume $\ell = 100$ ms. Then, according to \emph{Theorem~\ref{theorem:CACC+identical-gain-case}}, choose $k_a = 0.2$. We can obtain the lower bound of $h_w$ as  
 \begin{align}
  h_w > \frac{ 4 (\tau_0 + r k_a \ell) }{ ( r + 1 ) ( 1 + r k_a ) } = 0.35 \ \mbox{s}.     
 \end{align}
Choose $h_w = 0.4$ s, then the admissible region of $k_v$ and $k_p$ is provided in Fig.~\ref{fig:k_v_k_p_region_CACC_plus}. Then, choose $k_v = 0.16$, $k_p = 0.02$.      

\begin{figure}[!htb]
\centering{\includegraphics[scale=0.5]{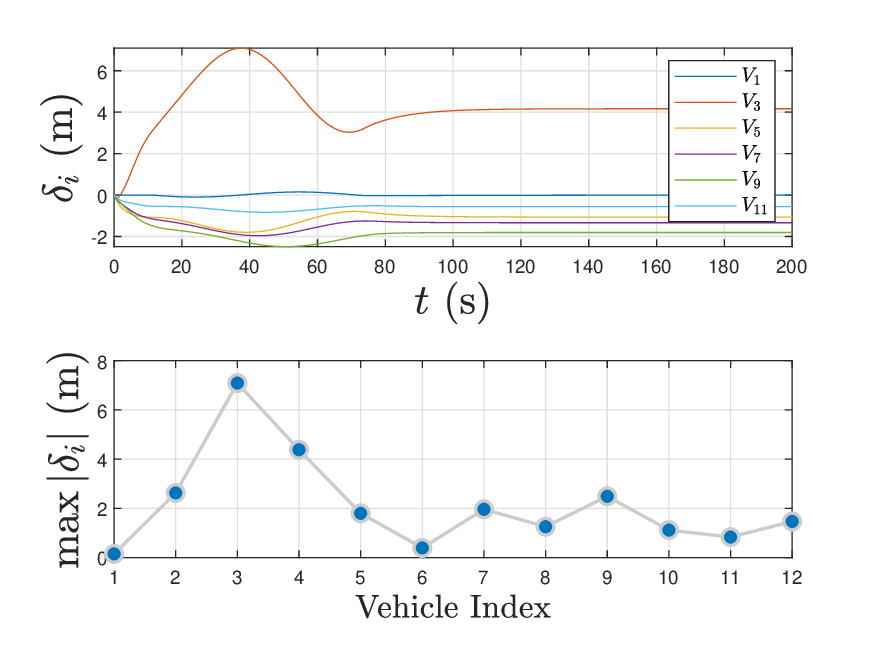}}
\caption{Inter-vehicular spacing errors (CACC+1, $r = 1$ and $d = 5$ m for vehicle 1, $r = 2$ and $d = 5$ m for vehicle 2, $r = 3$ and $d = 2.5$ m for vehicles 3 to 12).
\label{fig:CACC_plus_r1r2r3}}
\end{figure}

To address the heterogeneity of the platoon arising from the first $(r - 1)$ following vehicles with less than $r$ predecessors for communication, i.e., vehicles 1 and 2 when $r = 3$, one possible scheme is to let vehicle 1 adopt CACC with just lead vehicle information (i.e., $r = 1$ for vehicle 1), and vehicle 2 adopt the two predecessors following case of CACC+ (i.e., $r = 2$ for vehicle 2). When the control gains and time headway for vehicle 1 are chosen as per CACC, and for vehicle 2 as per CACC+ with $r = 2$ with control gains and time headway as $k_a = 0.2$, %$\tilde{k}_{a2} = 0.3$, 
$k_{v} = 0.35$, %$\tilde{k}_{v2} = 0.1$, 
$k_p = 0.03$, %$\tilde{k}_{p2} = 0.01$, 
$h_w = 0.6$ s, 
the evolution of the inter-vehicular spacing errors is depicted in Fig.~\ref{fig:CACC_plus_r1r2r3}. We refer to this case as CACC+1.  %\textcolor{red}{In Fig. 6 top graph, change the y-range to [-3,8].}{\color{blue}Response: The y-range has been changed as [-3, 8]. }   

\begin{figure}[!htb]
\centering{\includegraphics[scale=0.5]{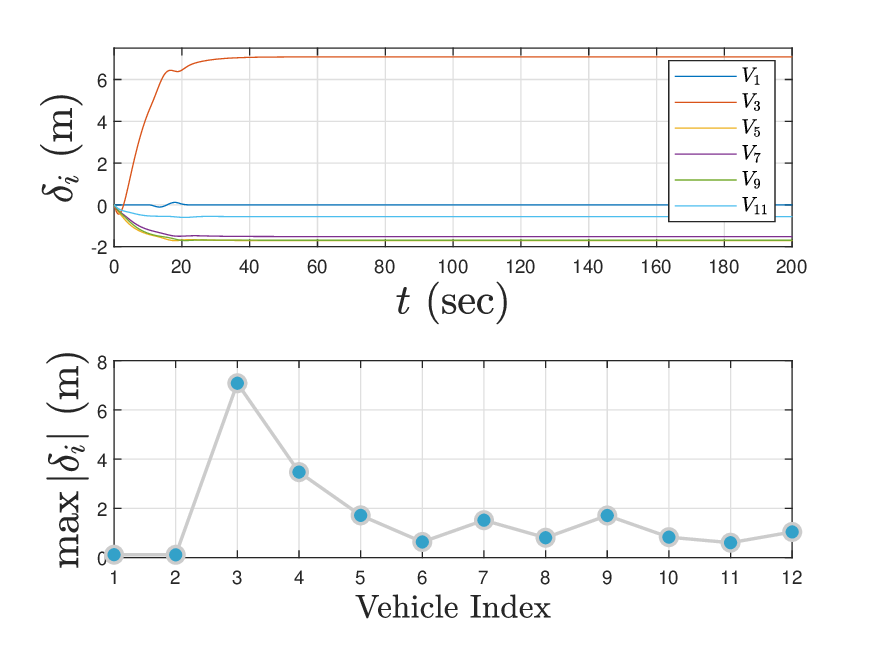}}
\caption{Inter-vehicular spacing errors (CACC+2, $r = 1$ and $d = 5$ m for vehicles 1 and 2, $r = 3$ and $d = 2.5$ m for vehicles 3 to 12).
\label{fig:MPF-delta-identical-hw}}
\end{figure}

Another possible option is to let the first two following vehicles adopt CACC, i.e., $r = 1$ for vehicles 1 and 2. 
Then, when the control gains and time headway for vehicles 1 and 2 are chosen as in the CACC case, 
the responses of the inter-vehicular spacing errors are provided in Fig.~\ref{fig:MPF-delta-identical-hw}. We refer to this case as CACC+2.

In addition, from Fig.~\ref{fig:CACC_plus_r1r2r3} and Fig.~\ref{fig:MPF-delta-identical-hw}, note that the inter-vehicular spacing errors of vehicles 3 to 12 are nonzero before the disturbance acts on the platoon, and maintain nonzero values at steady state; this may be due to the communication delay in velocity and position in addition to acceleration for CACC+, which causes (i) the control inputs for vehicles 3 to 12 to be nonzero before the disturbance acts on the platoon and (ii) the steady state values of the inter-vehicular spacing errors to be nonzero.  
Besides, it is worth noting that the amplification of the inter-vehicular spacing errors from vehicle 1 to vehicle 2 (and/or from vehicle 2 to vehicle 3) can propagate along the stream via communication and cause amplification between some adjacent vehicles. 

\begin{figure}[!htb]
\centering{\includegraphics[scale=0.5]{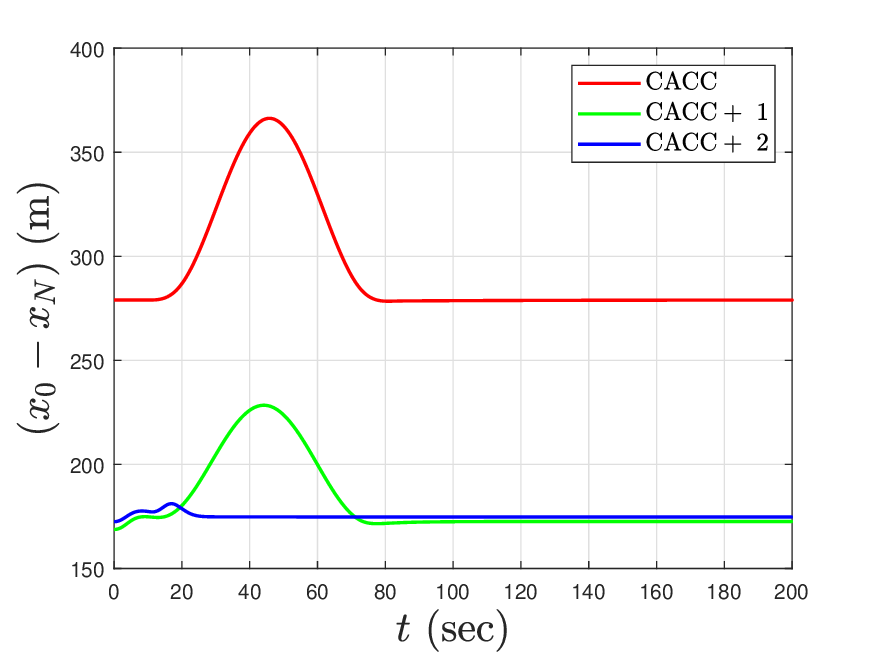}}
\caption{The length of the platoon (CACC+1 denotes the case where $r = 1$ for vehicle 1 and $r = 2$ for vehicle 2; CACC+ 2 denotes the case where $r = 1$ for both vehicle 1 and vehicle 2).
\label{fig:comparison_of_length}}
\end{figure}

Moreover, the comparison of the total length of the platoon during motion,  given by $x_0(t) - x_{N}(t)$, is provided in Fig.~\ref{fig:comparison_of_length}, from which it can be seen that the length of the platoon can be reduced significantly by using CACC+ compared to CACC, thus, increasing throughput. 

From the simulation results, it can be observed that for both the CACC and CACC+ cases, using the achieved design method, one can synthesize control gains and time headway to ensure robust string stability of the platoon. In addition, the length of the platoon can be substantially smaller by employing CACC+ than CACC.               

\section{Conclusion} \label{section:conclusion}
We have investigated the benefits of V2V communication in the presence of signal delay in communicated signals for connected and autonomous vehicles employing a constant time headway policy. We have derived a lower bound for time headway that is dependent on the latency in information communicated from the predecessor vehicles due to V2V communication, while ensuring that the CACC/CACC+ system is robustly string stable for parasitic lag and communication latency. We have provided a systematic analysis through which one can select control gains and time headway for the given latency. We have also provided numerical simulation results to corroborate the main results. For future work, we plan to investigate the effects of (i) quantization/bandwidth constraints in the design and study its effect on the lower bound of the time headway; and (ii) stochastic packet loss under sampled-data control. 
%%%%%%%%%%%%%%%%%%%%%%%%%%%%%%%%%%%%%%%%%%%%%%%%%%%%%%%%%%%%%%
\bibliographystyle{IEEEtran}

\bibliography{ref}

\end{document}